\documentclass[article]{mn2e}

\title[A Circumbinary Planet orbiting RR Cae]{A Circumbinary Planet in Orbit Around the Short-Period White-Dwarf Eclipsing Binary RR Cae}

\author[Qian, et al.]{Qian S.-B.$^{1,2,3}$, Liu L.$^{1,2,3}$, Zhu
L.-Y.$^{1,2}$, Dai Z.-B.$^{1,2}$, Fern\'{a}ndez Laj\'{u}s,
E.$^{4,5}$,
\newauthor{and Baume G. L.$^{4}$}\\
 $^1$National Astronomical Observatories/Yunnan Observatory, Chinese
     Academy of Sciences, P.O. Box 110,\\
     650011 Kunming, P. R. China (qsb@ynao.ac.cn)\\
 $^2$Key laboratory of the structure and evolution of celestial
  objects, Chinese Academy of Sciences, P.O. Box 110,\\ 650011
  Kunming, P. R. China\\
 $^3$Graduate University of the Chinese Academy of Sciences, 100049 Beijing, P. R. China
 Yuquan Road 19\#,\\
 Sijingshang Block, 100049 Beijing City, P. R. China\\
 $^4$Facultad de Ciencias Astron\'{o}micas y
Geof\'{i}sicas, Universidad Nacional de La Plata, 1900 La Plata,
Buenos Aires, Argentina\\
 $^5$Instituto de Astrofisica de La Plata (CCT La plata
- CONICET/UNLP), Argentina\\}

\date{}
\begin{document}
\maketitle

\begin{abstract}
By using six new determined mid-eclipse times together with those
collected from the literature, we found that the Observed-Calculated
(O-C) curve of RR Cae shows a cyclic change with a period of
11.9\,years and an amplitude of 14.3\,s, while it undergoes an
upward parabolic variation (revealing a long-term period increase at
a rate of $\dot{P}=+4.18(\pm0.20)\times{10^{-12}}$). The cyclic
change was analyzed for the light-travel time effect that arises
from the gravitational influence of a third companion. The mass of
the third body was determined to be
$M_3\sin{i}^{\prime}=4.2(\pm0.4)\,M_{Jup}$ suggesting that it is a
circumbinary giant planet when its orbital inclination is larger
than $17.6^{\circ}$. The orbital separation of the circumbinary
planet from the central eclipsing binary is about $5.3(\pm0.6)$\,AU.
The period increase is opposite to the changes caused by angular
momentum loss via magnetic braking or/and gravitational radiation,
nor can it be explained by the mass transfer between both components
because of its detached configuration. These indicate that the
observed upward parabolic change is only a part of a long-period
(longer than 26.3\,years) cyclic variation, which may reveal the
presence of another giant circumbinary planet in a wide orbit.

\end{abstract}


\section{Introduction}

RR Cae is one of a few double-lined eclipsing binaries (e.g., V471
Tau, QS Vir, RXJ2130.6+4710, and RR Cae) containing white dwarfs
where the masses and radii of both component stars can be determined
accurately (e.g., O'Brien et al., 2001; O'Donoghue et al., 2003;
Maxted et al., 2004, 2007). It was discovered to be an eclipsing
binary by Krzeminski (1984) in which the cool white dwarf component
is eclipsed by an M-type dwarf companion every 7.3\,h. The amplitude
of the radial velocity (of the dMe star) throughout the orbit was
mentioned as 370 km/s. Bruch \& Diaz (1998) published an R-band
light curve and classified the spectral type of the red-dwarf
companion as M5\,V or M6\,V. Further spectroscopy around the
$H_{\alpha}$ line was presented by Bruch (1999) who derived a
spectroscopic orbit for the M-dwarf and determined absolute
parameters of the binary system. The mass of the M-type dwarf was
estimated as 0.09\,$M_{\odot}$ by Bruch \& Diaz (1998) and Bruch
(1999), which was much lower than expected given its spectral type.
Several narrow absorption lines of neutral metals, e.g., Al\,I,
Fe\,I and Mg\,I, were found by Zuckerman et al. (2003) with Keck
telescope HIRES echelle observations, which were thought to have
been accreted onto the surface of the white dwarf from the M-dwarf.
A more detailed photometric and spectroscopic investigation was
recently presented by Maxted et al. (2007). They showed that RR Cae
is a detached binary containing a cool white dwarf primary with a
mass of $0.44\,M_{\odot}$ and an $M_4$-type secondary with a mass of
$0.182\,M_{\odot}$.

Searching for planetary companions to short-period white dwarf
binaries will shed light on the formation and evolution of planets,
as well as provide insight into the ultimate fate of planets and the
interaction between planets and evolved stars. Because of the very
small size of the white dwarfs, eclipse times of this type of binary
system can be determined with a high precision, and very
small-amplitude variations in the orbital periods could be detected
by analyzing the observed-calculated (O-C) diagrams. Therefore, they
are the most promising targets to search for circumbinary brown
dwarfs and planets by analyzing the light-travel-time effect. To
date, a few substellar companions to eclipsing white-dwarf binaries
were found by using this method, such as V471 Tau (e.g., Guinan \&
Ribas 2001), DP Leo (Qian et al. 2010a; Beuermann et al. 2011), QS
Vir (Qian et al. 2010b, Almeida \& Jablonski 2011), NN Ser (Qian et
al. 2009; Beuermann et al. 2011), and HU Aqr (Qian et al. 2011). In
this letter, by including 6 new mid-eclipse times, the variations of
the O-C curve of RR Cae were analyzed. Our results suggest that
there is a circumbinary giant planet orbiting the eclipsing binary,
and there is some evidence for a second, more distant companion.

\section{New observations and the changes of the O-C curve}

\begin{figure}
\centering
\vbox to4.0in{\rule{0pt}{5.5in}} \includegraphics{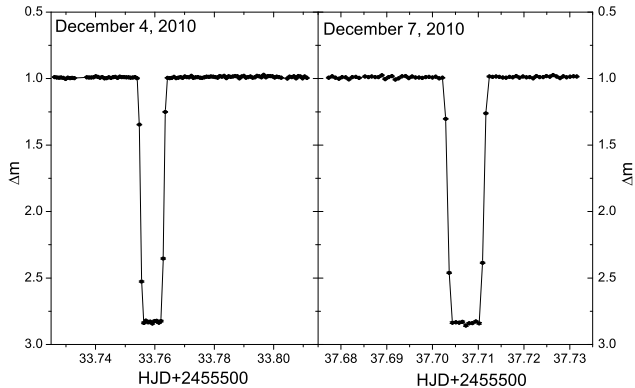}
\caption{Light curves of RR Cae in V band obtained by using the 2.15-m Jorge Sahade
telescope on December 4 and 7, 2010. The coordinates of the
comparison star are: $\alpha_{2000}=04^{h}20^{m}46.1^{s}$ and
$\delta_{2000}=-48^\circ41'46.8"$, while those of the check star
are: $\alpha_{2000}=04^{h}20^{m}41.2^{s}$ and
$\delta_{2000}=-48^\circ37'35.4"$.} \label{fig1}
\end{figure}

The white dwarf-red-dwarf eclipsing binary, RR Cae, was monitored on
December 4 and 7, 2010 with Roper Scientific, Versarray 1300B
camera, with a thinned EEV CCD36-40 de $1340\times{1300}$ pix CCD
chip, attached to the 2.15-m Jorge Sahade telescope at Complejo
Astronomico El Leoncito (CASLEO), San Juan, Argentina. During the
observation, V filter was used and an exposure time for each CCD
image was adopted as 50\,s. The clock of the control computer
operating the VersArray 1300B CCD camera is calibrated against UTC
time by the GPS receiver's clock. Two nearby stars that have similar
brightness in the same field of view of the telescope were chosen as
the comparison star and the check star, respectively. All images
were reduced by using PHOT (measure magnitudes for a list of stars)
of the aperture photometry package of IRAF. The corresponding light
curves are displayed in Fig. 1; the eclipse depth is about 1.84
magnitudes.

\begin{figure}
\centering
\vbox to4.0in{\rule{0pt}{5.5in}} \includegraphics{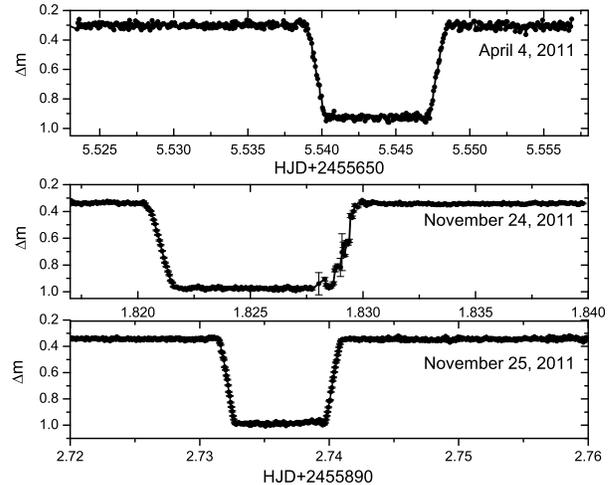}
\caption{Three
white-light eclipse profiles of the short-period eclipsing binary RR
Cae observed in April and December, 2011.} \label{fig2}
\end{figure}

\begin{table*}
\caption{New CCD times of light minimum of RR Cae.}
\begin{center}
\begin{tabular}{llll}\hline
HJD (days) & BJD (days)  & Errors (days) &  Filters\\\hline\hline
2455533.75855 & 2455533.75931 & $\pm0.00005$   & V    \\
2455537.70673 & 2455537.70749 & $\pm0.00005$   & V    \\
2455655.54371 & 2455655.54447 & $\pm0.00002$   & N    \\
2455889.69915 & 2455889.69991 & $\pm0.00002$   & N    \\
2455891.82509 & 2455891.82585 & $\pm0.00002$   & N    \\
2455892.73619 & 2455892.73695 & $\pm0.00002$   & N    \\
\hline\hline
\end{tabular}
\end{center}
\end{table*}

\begin{figure}
\centering
\vbox to4.0in{\rule{0pt}{5.5in}} \includegraphics{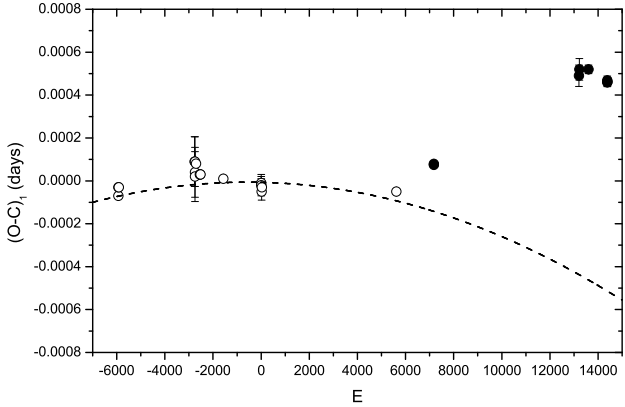}
\caption{Residuals
from the linear ephemeris of Maxted et al. (2007). Open circles
refers to the data published by Bruch \& Diaz (1998) and by Maxted
et al. (2007), while solid circles to the observations obtained by
Parsons et al. (2010) and the present authors. The dashed line
represents the quadratic fit derived by Maxted et al. (2007).}
\label{fig3}
\end{figure}

To get more mid-eclipse times of the binary star, we re-observed it
in April and November 2011. The exposure time for each CCD image was
adopted as 2\,s and no filters were used. Three eclipse profiles are
shown in Fig. 2. As displayed in the figure, the time of the
flat-bottomed minimum was estimated as 8.4\,minutes. By using the
midpoint times of steep ingress and egress of eclipse, we determined
six mid-eclipse times which are listed in Table 1. Since Barycentric
Dynamical Time (BJD) is a precise time system, during the analyzing
we applied this time system. The HJD times were converted to BJD
ones with the code of Stumpff (1980) and those mid-eclipse times in
both the HJD and BJD are shown in Table 1.

Times of mid-eclipse of RR Cae were published by a few authors
(e.g., Krzeminski, 1984; Bruch \& Diaz, 1998; Maxted et al., 2007;
Parsons et al., 2010). The $(O-C)_1$ values of all available
mid-eclipse times were calculated with the linear ephemeris derived
by Maxted et al. (2007),
\begin{equation}
Min.I= BJD\,2451523.048567+0^{d}.3037036366\times{E},
\end{equation}
where BJD\,2451523.048567 is the initial epoch and 0.3037036366 days
is the orbital period. The corresponding $(O-C)_1$ diagrams are
shown in Figs. 3 and 4 along with the epoch number E. By removing
the mid-eclipse time, BJD\,2445927.91665, determined by Krzeminski
(1984), Maxted et al. (2007) obtained the following quadratic
ephemeris,
\begin{eqnarray}
Min.I&=&BJD\,2451523.048560+0^{d}.3037036340\nonumber\\
  & &-2.27\times{10^{-12}}\times{E^{2}},
\end{eqnarray}
with their mid-eclipse times and those published by Bruch \& Diaz
(1998). The quadratic term in this ephemeris reveals a period
decrease as a rate of $\dot{P}/P=-5\times{10^{-12}}$. However, as
shown in Fig. 3, our data and those determined by Parsons et al.
(2010) do not follow the general trend predicted by the quadratic
ephemeris suggesting that the variation of the $(O-C)_1$ curve of RR
Cae is very complex.

\begin{figure}
\centering
\vbox to4.0in{\rule{0pt}{5.5in}} \includegraphics{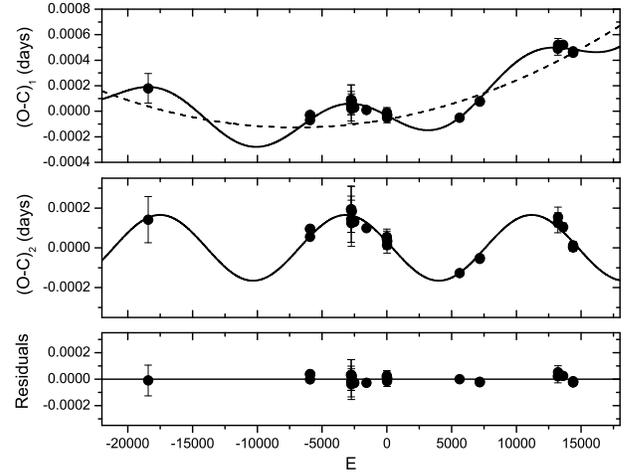}
\caption{A plot of
the $(O-C)_1$ curve of RR Cae from the linear ephemeris of Maxted et
al. (2007) is shown in the upper panel. The solid line in the panel
refers to a combination of an upward parabolic variation and a
cyclic change. The dashed line represents the upward parabolic
variation that reveals a continuous increase in the orbital period.
The $(O-C)_2$ values with respect to the quadratic part of Eq. (3)
are displayed in the middle panel where a cyclic change is more
clearly seen. After both the upward parabolic change and the cyclic
variation were removed, the residuals are plotted in the lowest
panel.} \label{fig4}
\end{figure}

To fit the $(O-C)_1$ curve satisfactorily, a combination of an
upward parabolic variation and a cyclic change are required (solid
line in the upper panel in Fig. 4). A least-square solution of all
available data leads to,
\begin{eqnarray}
(O-C)_1&=&-0.000064(\pm0.000005)\nonumber\\
&&+1.79(\pm0.08)\times{10^{-8}}\times{E}\nonumber\\
&&+1.27(\pm0.06)\times{10^{-12}}\times{E^{2}}\nonumber\\
&&+0.000165(\pm0.000013)sin[0.^{\circ}02590\nonumber\\
&&(\pm0.^{\circ}00010)\times{E}+ 169.2^{\circ}(\pm0.8^{\circ})],
\end{eqnarray}
which suggests a cyclic oscillation with a very small amplitude of
14.3\,s and a period of 11.9\,years. The quadratic term in Eq. (3)
indicates a linear increase at a rate of
$\dot{P}=+4.18(\pm0.20)\times{10^{-12}}$\,s/s (or 1.3\,s in about
10000\,years). The dashed line in the upper panel of Fig. 4 refers
to the linear period increase, while the solid one represents the
combination of the linear increase and the cyclic change. The
$(O-C)_{2}$ values with respect to the quadratic part Eq. (3) are
displayed in the middle panel where a small-amplitude periodic
variation can be seen more clearly. After both of the linear
increase and the cyclic change were subtracted, the residuals are
plotted in the lowest panel where no variations can be traced
indicating that Eq. (3) describes the general trend of the $(O-C)_1$
curve well.

\section{Discussions and conclusions}

As shown in the middle panel of Fig. 4, after the upward parabolic
variation is subtracted from the $(O-C)_1$ curve, the $(O-C)_2$
residuals suggest that there is a cyclic variation with a period of
11.9\,years. The cyclic change of the mid-eclipse times represent
either true or apparent variations (e.g., caused by apsidal motion
or the presence of a third body) in the orbital period. As for the
true cyclic variation, it was usually attributed to the mechanism of
Applegate (1992). In this mechanism, the changes in the internal
constitution of the cool component star during the magnetic activity
cycles result in the variation of the orbital period through
spin-orbit coupling. However, the secondary of RR Cae is a fully
convective M4-type star; Applegate's mechanism in such cool stars is
generally too feeble to explain the observed amplitudes (e.g.,
Brinkworth et al. 2006). On the other hand, RR Cae is a short-period
binary with an orbital period of 7.3\,h, and the strong tidal
interaction is expected to circularize the orbit efficiently.
Apsidal motion cannot account for the cyclic variation in the
mid-eclipse times. Therefore, the periodic change of the $(O-C)_2$
residuals can be plausibly explained as light-travel-time effect
caused by the motion of the eclipsing binary via the presence of an
invisible companion.

The sine-like variation of the $(O-C)_2$ curve in the middle of Fig.
4 suggests that the eccentricity of the orbit of the tertiary
component is close to zero. The projected radius of the orbit of the
eclipsing pair rotating around the barycenter of the triple system
was computed with this equation,
\begin{equation}
a_{12}^{\prime}\sin{i}^{\prime}=A_3\times{c},
\end{equation}
where $A_3$ is the amplitude of the O-C oscillation and $c$ is the
speed of light, i.e.,
$a_{12}^{\prime}\sin{i}^{\prime}=0.029(\pm0.002)$\,AU (1 AU is the
mean distance between the earth and the Sun). Then, by using the
absolute parameters determined by Maxted et al. (2007), a
calculation with the following equation,
\begin{equation}
f(m)=\frac{4\pi^{2}}{GP_{3}^{2}}\times{(a_{12}^{\prime}\sin{i}^{\prime})}^{3}=\frac{(M_{3}\sin{i^{\prime}})^{3}}{(M_{1}+M_{2}+M_{3})^{2}},
\end{equation}
where $G$ and $P_3$ are the gravitational constant and the period of
the $(O-C)_2$ oscillation, yields the mass function and the mass of
the tertiary companion as:
$f(m)=1.7(\pm0.3)\times{10^{-7}}\,M_{\odot}$ and
$M_3\sin{i}^{\prime}=0.00401(\pm0.00037)\,M_{\odot}$=$4.2(\pm0.4)$
$M_{Jup}$, respectively. If the orbital inclination of the third
body is larger than $17.6^{\circ}$, the mass of the tertiary
component corresponds to: ${M_3}\le0.014\,M_{\odot}$, and it should
be an extrasolar planet. Therefore, with 95.3\% probability, the
third body is a giant planet (by assuming a random distribution of
orbital plane inclination). The parameters of the circumbinary
giant planet are shown in Table 2. When the orbital inclination
equals $90^{\circ}$, the orbital distance between the circumbinary
planet and the central binary is about $5.3(\pm0.6)$\,AU.

It is interesting to point out that the proposed planet has a period
of 11.9 years and an eccentricity consistent with zero, which are
essentially identical to the period and eccentricity of Jupiter
($P=11.86$\,years, $e=0.05$). This raises the question as to
whether the gravitational influence of Jupiter has been properly
accounted for in the timing calculations. During the analysis, we
used Barycentric Dynamical Time, which considers the influence of
all planets. The light-travel time amplitude induced by the orbit of
Jupiter about the Solar System barycentre is 2.48 seconds. Thus, an
error omitting Jupiter's influence cannot have caused the observed
cyclic signal with an amplitude of 14.3 seconds.

\begin{table}
\caption{The derived planetary parameters in RR Cae.}
\begin{center}
\begin{tabular}{ll}
\hline
Parameters & Value and uncertainty\\
\hline\hline
Period ($P_3$) & 11.9$\pm$0.1yr   \\
Eccentricity ($e_3$) & 0.0    \\
Amplitude ($K_3$) & 14.3$\pm$1.1s \\
$f(m)$ & $1.7(\pm0.3)\times{10^{-7}}\,M_{\odot}$\\
$M_3{\sin{i}^{\prime}}$ & 4.2$\pm$0.4$M_{Jup}$\\
$a_3$($\sin{i}^{\prime}=90^{\circ}$) & $5.3(\pm0.6)$\,AU\\
$\chi^{2}$ & 1.03 \\
RMS of the least square fit & 3.05$\times$10$^{-5}$days\\
\hline\hline
\end{tabular}
\end{center}
\end{table}

The upward parabolic variation in the $(O-C)_1$ curve indicates that
the period of RR Cae is increasing continuously at a rate of
$\dot{P}=+4.18(\pm0.20)\times{10^{-12}}$\,s/s. Since RR Cae is a
detached binary system where both components are in the critical
Roche lobes, no lobe-filling mass transfers occur between both
components. Moreover, the angular momentum loss caused by
gravitational radiation or/and magnetic braking should produce a
decrease in the orbital period. Therefore, neither mass transfer nor
angular momentum loss can explain the period increase of RR Cae. All
of these suggest that the observed upward parabola in the $(O-C)_1$
curve may be an apparent variation in the orbital period due to the
influence of an additional orbiting body. As in the cases of the
eclipsing polar HU Aqr (Qian et al. 2011) and the sdB-type
eclipsing binary NY Vir (Qian et al. 2012), it may be only a part of
a long-period (longer than 26.3\,years) cyclic change, possibly
revealing the presence of another circumbinary planet in a wider
orbit. Some authors (e.g., Horner et al. 2011; Wittenmyer et al.
2012) proposed that the serious problem for the presence of the
fourth body is the dynamical stability of the circumbinary planetary
system. However, dynamical simulations by Hinse et al. (2011)
suggested a family of stable orbital solutions, though for shorter
timescales than tested by Horner et al. (2011).

If the cyclic variations in the O-C curve of RR Cae are really
caused by the light-travel time effects via the presence of
circumbinary planets, they should be strictly periodic. To check the
existence of the planetary system, more mid-eclipse times are needed
in the future. Moreover, as pointed out by Qian et al. (2012), at
the maximum points of the cyclic O-C changes, RR Cae should be at
the farthest position of the orbit, while the circumbinary planets
are closest to the observer. If the tertiary companion is coplanar
to the central eclipsing binary, the circumbinary planet should be
in the light of the binary system and transit the binary component
stars. Therefore, searching for the transits of the binary
components by the circumbinary planets at the O-C maxima can
ascertain the presence of the planetary system. However, the
possibility of transits is very low ($0.026\%$ by assuming a radius
of $0.3\,R_{\odot}$ for the M dwarf stars). Recently, a Saturn-like
planet transiting the M-type components in the eclipsing binary
Kepler-16 was reported by Doyle et al. (2011), which provides the
first direct evidence for the presence of a circumbinary planet.

\section*{acknowledgments}
This work is partly supported by Chinese Natural Science Foundation
(No.11133007, No.10973037, No.10903026, and No.11003040) and by the
West Light Foundation of the Chinese Academy of Sciences. New CCD
photometric observations of RR Cae were obtained with the 2.15-m
"Jorge Sahade" telescope.

\end{document}